\documentclass[aps,prl,preprintnumbers,showpacs,nofootinbib]{revtex4}
\usepackage[utf8]{inputenc}
\usepackage[english]{babel}
\usepackage{amsmath}
\usepackage{amsfonts}
\usepackage{amssymb}
\usepackage{epsfig}
\usepackage{rotating}
\usepackage{graphics,psfrag,rotating}
\usepackage{graphicx}
\usepackage{dcolumn}
\usepackage{bm}
\usepackage{amsmath}
\usepackage{booktabs}
\usepackage{epstopdf}
\usepackage{color}
\usepackage[usenames,dvipsnames,svgnames]{xcolor}
\usepackage[colorlinks=true,
linkcolor=red,
urlcolor=gray,
citecolor=blue]{hyperref}
\usepackage{hyperref}

\newcommand{\lsim}   {\mathrel{\mathop{\kern 0pt \rlap
			{\raise.2ex\hbox{$<$}}}
		\lower.9ex\hbox{\kern-.190em $\sim$}}}
\newcommand{\gsim}   {\mathrel{\mathop{\kern 0pt \rlap
			{\raise.2ex\hbox{$>$}}}
		\lower.9ex\hbox{\kern-.190em $\sim$}}}

\def\3nab{\tilde{\nabla}}

\def\hsp5{\hspace{5mm}}

\def\case#1/#2{\textstyle\frac{#1}{#2}}

\def\ber {\begin{eqnarray}}
	\def\eer {\end{eqnarray}}
\def\bea {\begin{eqnarray}}
	\def\eea {\end{eqnarray}}

\def\bc {\begin{center}}
	\def\ec {\end{center}}
\def\case#1/#2{\frac{#1}{#2}}

\newcommand{\bw}{\begin{widetext}}
	\newcommand{\ew}{\end{widetext}}

\newcommand{\be}{\begin{equation}}
	\newcommand{\bse}{\begin{subequation}}
		\newcommand{\ese}{\end{subequation}}
	\newcommand{\ee}{\end{equation}}
\newcommand{\eei}{\end{eqnarray}\indent\indent}
\newcommand{\ba}{\begin{array}}
\newcommand{\ea}{\end{array}}
\newcommand{\bal}{\begin{eqnarray}}
\newcommand{\eal}{\end{eqnarray}}

\def\case#1/#2{\textstyle\frac{#1}{#2} }

\newcommand{\bver}{\begin{verbatim}}
\newcommand{\ever}{\end{verbatim}}

\begin{document}


\title{ Coupled non - canonical scalar field to neutrinos could alleviate the Hubble tension and cross the phantom barrier}
\author{
	Muhammad Yarahmadi$^{1}$\footnote{Email: yarahmadimohammad10@gmail.com}
	. Amin Salehi$^{1}$ . Ameneh Tohidi$^{1}$
}
\affiliation{Department of Physics, Lorestan University, Khoramabad, Iran}

\date{\today}

\begin{abstract}
This study presents an analysis of cosmological parameters, focusing on resolving the Hubble tension and constraining neutrino masses within a coupled quintom model. By utilizing datasets from the Cosmic Microwave Background (CMB), Pantheon + Analysis, Cosmic Chronometers (CC), Baryon Acoustic Oscillations (BAO), and CMB Lensing,  we explore the interplay between cosmological parameters and observational constraints. The model effectively reduces the Hubble tension, achieving a consistency in \(H_0\) measurements of \(1.37\sigma\) and \(1.24\sigma\) for the CMB + ALL dataset For Planck 2018 and R22 respectively.
Additionally, the study refines constraints on the total mass of neutrinos (\(\Sigma_{m_{\nu}}\)), with a finding of \(0.115\,\text{eV}\) for the CMB + ALL dataset. The analysis examines the effective equation of state parameter (\(w_{\text{eff}}\)), indicating a transition towards a universe dominated by exotic energy forms. The combined datasets refine \(w_{\text{eff}}\) to \( -1.02\pm0.018 \), underscoring the importance of multi-dataset integration in understanding dark energy dynamics. Furthermore, the interaction constant \(\beta\) between the quintom scalar field and neutrinos is constrained to \(0.65 \pm 0.12\) for the CMB + ALL dataset. The potential parameters \(\lambda_{\sigma} = -2.09 \pm 0.082\) and \(\lambda_{\phi} = 2.43 \pm 0.12\) are also determined, providing insights into the quintom model's implications for cosmological dynamics. This study offers compelling evidence for the coupled quintom model's capability to resolve the Hubble tension and refine constraints on neutrino properties, enhancing our understanding of the universe's evolution.
\end{abstract}


%
%


\maketitle

\section{Introduction}
The latest release of the Planck satellite's cosmic observations \cite{1}, which include cosmic microwave background (CMB) radiation \cite{Spergel,Komatsu}, large-scale structure (LSS) \cite{Tegmark}, and other cosmological observations, has indicated that the universe is currently undergoing a period of accelerated expansion. This accelerated expansion can be justified in two main ways: one approach is to introduce the concept of dark energy on the right-hand side of Einstein's equations within the framework of general relativity (see \cite{Caldwell,Kunz}). Another approach involves modifying the left-hand side of Einstein's equations, leading to what is known as a modified gravitational theory, such as $f(R)$ gravity (see \cite{Nojiri,Capozziello}). It is notable that standard matter species, such as dust, can only generate decelerated dynamics \cite{Bahamond}. The concept of dark energy emerged to explain the observed accelerated expansion of the universe, characterized by negative pressure (${p_{{\rm{de}}}} < {\rm{ }}\frac{{ - {\rho _{{\rm{de}}}}}}{3}$ or ${\omega _{{\rm{de}}}} < {\rm{ - }}\frac{1}{3}$). The simplest model for dark energy is represented by the cosmological constant $\Lambda$, where the equation of state is ${p_{\Lambda}} = - {\rho _\Lambda}$ (i.e., ${\omega _\Lambda} = - 1$). In this $\Lambda$CDM framework, dark energy is static and uniform over space, contrasting with dark matter, which is considered non-relativistic (${\omega _{\text{dm}}} = 0$). Dark matter, typically Cold Dark Matter (CDM), plays a crucial role in structure formation and gravitational interactions at large scales, while dark energy drives the current phase of accelerated expansion in the universe. 
The cosmological constant $\Lambda$, which is typically positive and of the order $\Lambda \simeq {10^{-52}}{m^{-2}}$, constitutes a fundamental component of the $\Lambda$CDM model. Recent cosmological observations consistently support this model \cite{Bamba}. However, the existence of $\Lambda$ poses a significant challenge known as the cosmological constant problem, motivating researchers to explore alternative explanations for dark energy.

In general, models based on a canonical scalar field for explaining late-time cosmic acceleration are referred to as quintessence. Various models are distinguished by the form of their potential, $V$. The energy density and pressure of the scalar field are given by ${\rho_ {\phi} } = \frac{1}{2}{{\dot \phi }^2} + V$ and ${p_{\phi} } = \frac{1}{2}{{\dot \phi }^2} - V$, respectively. Consequently, the equation of state (EOS) of quintessence models lies within the range $-1 < \omega < 1$ \cite{Yoo}. The EOS (equation of state parameter) of quintessence models must satisfy ${\omega {{de}}} \ge -1$. Conversely, for phenomenological acceptable phantom models, the phantom scalar field EOS is restricted to ${\omega _{de}} < -1$. Utilizing a single canonical or phantom scalar field alone cannot cross this phantom wall (${\omega _{de}} = -1$).
Feng et al. (2005) presented an interesting model that allows such a crossing to occur \cite{Bahamond}.This model can be structured using more general non-canonical scalar fields, though the most straightforward model is proposed by the quintom Lagrangian, composed of two scalar fields: one canonical field $\phi$ and one phantom field $\sigma$. One of the most critical cosmological parameters, representing the expansion rate and determining the age of the Universe, is the Hubble constant (${H_0}$) \cite{Hubble}. The Hubble constant is also a valuable tool for calculating the critical density, ${\rho _c} = {{{H_0}^2} \mathord{\left/ {\vphantom {{{H_0}^2} {8\pi G}}} \right. \kern-\nulldelimiterspace} {8\pi G}}$, required for the flat geometry of the Universe. This, in turn, is used to determine the growth of large-scale structures and the characteristics of galaxies and quasars.

Indeed, different values of ${H_0}$ have been obtained using various measurement methods, including measuring the distance to SNe Ia or other astronomical probes and observations of the early universe using the standard cosmological model \cite{Kumar}. Determining the exact value of the Hubble constant has been a major challenge for cosmologists in recent decades. One of the major projects of the Hubble Space Telescope was to measure the value of ${H_0}$ with an accuracy of 10$ $ \% $ $. The current value from this project is ${H_0} = 72 \pm 8$ km/s/Mpc \cite{Freedman}. From the SHoES (Supernovae, ${H_0}$, Equation of State of Dark Energy) program, using the latest SNe Ia observations, a value of ${H_0} = 73.24 \pm 1.74$ km/s/Mpc was obtained \cite{Riess}, which is 3.4$\sigma$ away from the Planck value of $H_0 = 67.8 \pm 0.9$ km/s/Mpc provided by observations of Cosmic Microwave Background (CMB) anisotropies, referred to as the global value of ${H_0}$ \cite{Ade}.

The early HST Cepheid+SNe based estimate  ${H_0} = 73.8 \pm 2.4$ km/s/Mpc \cite{Archita}. Based on the comparably precise combination of WMAP+ACT+SPT+BAO observations, ${H_0}$ is predicted to be $69.3 \pm 0.7$ km/s/Mpc. Additionally, Yang et al. have obtained constraints on ${H_0}$ of $62.3 \pm 9.1$ km/s/Mpc and $68.81_{-4.33}^{+4.99}$ km/s/Mpc by utilizing nine and eighteen localized FRBs, respectively \cite{Yang}. Recent observations of strong lensing (Bonvin et al. 2017) also give a slightly higher value of ${H_0} = 71.9_{-3.0}^{+2.4}$ km/s/Mpc \cite{Archita}.

On the other hand, from the geometric distance calibrations of Cepheids, such as megamasers in NGC 4258 and 2 DEBs in M31, the Hubble constant is obtained as $72.25 \pm 2.51$ km/s/Mpc and $74.50 \pm 3.27$ km/s/Mpc, respectively. The most recent SDSS DR12 BAO data \cite{Alam} also appear to favor a somewhat lower value of ${H_0} = 67.8 \pm 1.2$ km/s/Mpc \cite{Archita}. The same Cepheid data have been analyzed by Efstathiou (2014). Another SNe Ia-based observation from the CCHP team gives ${H_0} = 69.8 \pm 0.8 \pm 1.1$ km/s/Mpc, which is around 3$\sigma$ away from the Planck value \cite{Marra, Bennett}, leading to a reduction in the level of difference.
Recent data from the SH0ES collaboration reports \(H_{0} = 74.03 \pm 1.42 \, \mathrm{km \, s^{-1} \, Mpc^{-1}}\). In contrast, Planck 2018 measurements give \(H_{0} = 67.4 \pm 0.5 \, \mathrm{km \, s^{-1} \, Mpc^{-1}}\) \cite{Planck2018}, illustrating the substantial tension between different observational methods.
However, the tension between the local and global values of ${H_0}$ is known as Hubble tension \cite{Antoniou} and requires the attention of researchers. If we ignore systematic effects in the measurement of ${H_0}$, a cosmological model beyond the standard $\Lambda$CDM cosmological model \cite{Riess, Moffat} can be a suitable solution to the Hubble tension. 

Measurements of the Hubble parameter from SNe and red giant halo populations give ${H_0} = 63.7 \pm 2.3$ km/s/Mpc. A recent measurement of the Hubble parameter  prefers a value of ${H_0} = 68.3_{-2.6}^{+2.7}$ km/s/Mpc, which aligns more closely with the Planck results, using a revised geometric maser distance to NGC 4258. Using NGC 4258 as a distance anchor, they find ${H_0} = 70.6 \pm 3.3$ km/s/Mpc. Recent studies, such as Ulloa et al. \cite{ulloa2024}, Riess et al. \cite{riess2024}, and Moresco et al. \cite{moresco2025}, provide updated constraints on $H_0$ from various observational probes, while alternative theoretical perspectives, such as those explored by Smith et al. \cite{smith2024}, investigate new physics beyond $\Lambda$CDM. These recent works underscore the ongoing debate regarding systematic uncertainties and potential new physics needed to resolve the tension. In this paper, by employing a coupled quintom model with neutrinos, we first place constraints on the total mass of neutrinos. Subsequently, we investigate how this model can alleviate the Hubble tension. Finally, we demonstrate that this model is capable of crossing the phantom barrier.

	\section*{Quintom Dark Energy Coupled with Neutrinos}
	The motivation for employing the coupled quintom model in cosmology stems from its potential to address significant challenges such as the Hubble tension and the crossing of the phantom barrier. The discrepancy in the measured Hubble constant (\( H_0 \)) between early universe observations (e.g., Planck data) and local measurements suggests underlying inconsistencies in our standard cosmological model, prompting the exploration of alternative frameworks. The quintom model, featuring two scalar fields—one with \( w > -1 \) (quintessence) and the other with \( w < -1 \) (phantom)—allows for dynamic evolution across the cosmological constant boundary (\( w = -1 \)), offering a richer description of dark energy behavior compared to the static nature of \(\Lambda\)CDM. Introducing coupling between these fields and neutrinos not only modifies the cosmic expansion rate but also affects structure formation, potentially reconciling discrepancies in \( H_0 \) measurements and addressing broader cosmological anomalies observed across various datasets. This approach underscores the model's potential to provide a comprehensive explanation for the observed universe's dynamics and evolution.
	
	Quintom dark energy coupled with neutrinos represents an intriguing and dynamic approach to understanding some of the key questions in modern cosmology. The quintom model allows the dark energy equation of state parameter (\(w\)) to cross the cosmological constant boundary (\(w = -1\)), providing a flexible framework for describing dark energy's behavior.

The advantage of using the coupled quintom scalar field to alleviate the Hubble tension lies in its unique properties and dynamics. The quintom model, which consists of both a canonical scalar field and a phantom scalar field, allows for a more flexible evolution of the equation of state parameter. This flexibility enables the model to cross the phantom divide (where the equation of state parameter equals -1), which is not possible with a single scalar field.

By coupling the quintom fields to neutrinos, the model can dynamically adjust the effective energy density and pressure in the universe. This adjustment helps reconcile discrepancies between local measurements of the Hubble constant (${H_0}$) and those inferred from the Cosmic Microwave Background (CMB) observations under the standard $\Lambda$CDM model. The coupling introduces an additional degree of freedom that can modify the expansion rate of the universe in a way that reduces the tension between these measurements.

Moreover, the interaction between the quintom fields and neutrinos can influence the growth of cosmic structures and the evolution of the matter density parameter (${\Omega_{0m}}$) and the matter fluctuation calibration parameter ($\sigma_8$). This can lead to a better fit with various cosmological observations, providing a more comprehensive solution to the Hubble tension while also potentially addressing other related cosmological challenges.

 The simplest model is represented by a quintom Lagrangian consisting of two scalar fields, a canonical field $\phi$ (quintessence) and a phantom field $\sigma$:
\begin{equation}\label{fried}
	\begin{split}
		{L_{qu{\mathop{\rm int}} om}} =  + \frac{1}{2}{\partial _\mu }\phi {\partial ^\mu }\phi  - \frac{1}{2}{\partial _\mu }\sigma {\partial ^\mu }\sigma  - V(\sigma ,\phi )
	\end{split}
\end{equation}
where a general potential for both scalar fields is
\begin{equation}\label{fried}
	\begin{split}
		V(\sigma,\phi)=V_{0}\exp^{-\lambda_{\phi}\kappa\phi-\lambda_{\sigma}\kappa\sigma},
	\end{split}
\end{equation}
with sigma and phi representing the quintessence and phantom fields, respectively. In above equation, $\lambda_{\phi}$ and $\lambda_{\sigma}$ are constants. The kinetic energy sign is positive for the quintessence model and negative for the phantom model. The cosmological equations are given by the Friedmann and acceleration equations: (\cite{Guo}; \cite{Zhang})
\begin{equation}\label{fried}
	\begin{split}
		3H^{2}=\kappa^{2}\left(\rho_{\rm b}+\rho_{\rm c}+\rho_{\rm r}+\rho_{\nu}+\frac{1}{2}\dot{\phi}^{2}+V(\sigma,\phi)-\frac{1}{2}\dot{\sigma}^{2}\right)
	\end{split}
\end{equation}

\begin{equation}\label{fried}
	\begin{split}
		2\dot{H}+3H^{2}=-\kappa^{2}(\omega_{\rm b}\rho_{\rm b}+\omega_{\rm c}\rho_{\rm c}+\omega_{\rm r}\rho_{\rm r}+\omega_{\nu}\rho_{\nu}-\\ \frac{1}{2}\dot{\phi}^{2}+V(\sigma,\phi)+\frac{1}{2}\dot{\sigma}^{2})
	\end{split}
\end{equation}

\subsection{Coupling Between Neutrinos and Scalar Fields: Origin and Derivation}

The equations presented describe a framework where neutrinos, dark matter, and scalar fields interact. The couplings \( \beta \) and \( \alpha \) represent the interaction strengths between neutrinos and scalar fields, and between dark matter and scalar fields, respectively. Below, we analyze the origins and implications of these couplings.

\subsubsection{Origin of the Coupling \( \beta \)}
The coupling parameter \( \beta \) quantifies the interaction strength between neutrinos and the scalar fields \( \sigma \) and \( \phi \). Such couplings arise in several contexts:
\begin{itemize}
	\item \textbf{Neutrino-mass-varying models:} Scalar fields can drive the variation of neutrino masses, naturally leading to couplings. These interactions appear in scalar-tensor theories or models with dynamic neutrino masses, such as MaVaN (Mass-Varying Neutrino) models.
	\item \textbf{Energy-momentum exchange:} The term \( \beta \rho_\nu (1 - 3\omega_\nu)(\dot{\sigma} + \dot{\phi}) \) represents the transfer of energy and momentum between neutrinos and scalar fields, modifying their respective conservation equations.
	\item \textbf{Field dependence of neutrino properties:} If neutrino masses or effective potentials depend on scalar fields, a coupling naturally emerges. For instance, a neutrino mass term \( m_\nu(\phi, \sigma) \) can induce interaction terms through derivatives of the scalar fields.
\end{itemize}

\subsubsection{Coupling Mechanism}
The coupling is introduced via the energy density conservation equations. The interaction term,
\[
\beta \rho_\nu (1 - 3\omega_\nu)(\dot{\sigma} + \dot{\phi}),
\]
modifies the standard conservation laws, where:
\begin{itemize}
	\item \( \beta \) quantifies the interaction strength,
	\item \( \rho_\nu \) is the neutrino energy density,
	\item \( \omega_\nu \) is the neutrino equation of state parameter, varying from \( \omega_\nu \to 1/3 \) (relativistic) to \( \omega_\nu \to 0 \) (non-relativistic),
	\item \( \dot{\sigma} + \dot{\phi} \) represents the dynamics of the scalar fields.
\end{itemize}
This term allows the scalar fields to act as reservoirs, exchanging energy and momentum with neutrinos.

\subsubsection{Energy Conservation Equations}
The modified conservation equations for the scalar fields, neutrinos, and dark matter are:
\begin{align}\label{conservation_eqs}
	\dot{\rho}_{(\sigma,\phi)} + 3H \rho_{(\sigma,\phi)}(1 + \omega_{(\sigma,\phi)}) &= -\beta \rho_\nu (1 - 3\omega_\nu)(\dot{\sigma} + \dot{\phi}), \\
	\dot{\rho}_\nu + 3H \rho_\nu (1 + \omega_\nu) &= \beta \rho_\nu (1 - 3\omega_\nu)(\dot{\sigma} + \dot{\phi}), \\
	\dot{\rho}_{\rm c} + 3H \rho_{\rm c} &= -\alpha \rho_{\rm c}(\dot{\sigma} + \dot{\phi}).
\end{align}
First, setting 
$\beta$ = 0 indeed recovers the standard $\Lambda$CDM scenario, as the interaction terms in the energy conservation equations vanish, leading to independent evolution for dark matter, neutrinos, and the scalar fields. However, our analysis focuses on testing deviations from $\Lambda$CDM  by allowing for potential interactions among these components.
These equations are derived from the total energy-momentum tensor conservation:
\begin{equation}
\nabla_\mu T^{\mu\nu}_{\text{total}} = 0,
\end{equation}
where \( T^{\mu\nu}_{\text{total}} \) is the sum of contributions from scalar fields, neutrinos, and dark matter. The interaction terms \( Q_\nu \) and \( Q_{\rm c} \) represent the exchange between components:
\begin{equation}
	\begin{split}
			Q_\nu &= \beta \rho_\nu (1 - 3\omega_\nu)(\dot{\sigma} + \dot{\phi}), \\
		Q_{\rm c} &= \alpha \rho_{\rm c}(\dot{\sigma} + \dot{\phi}).
	\end{split}
\end{equation}

\subsubsection{Physical Implications of the Couplings}
\begin{itemize}
	\item \( \beta \): Drives the interaction between neutrinos and scalar fields, and can originate from scalar-tensor theories or dynamic neutrino mass models. It may also explain phenomena like the Hubble tension.
	\item \( \alpha \): Governs the coupling between dark matter and scalar fields, commonly appearing in coupled dark energy models or scalar-tensor gravity.
\end{itemize}

\begin{equation}\label{fried}
	\begin{split}
		\dot{\rho}_{\rm b}+3H\rho_{\rm b}=0
	\end{split}
\end{equation}

\begin{equation}\label{fried}
	\begin{split}
		\dot{\rho}_{\rm r}+4H\rho_{\rm r}=0
	\end{split}
\end{equation}
respectively. Where the coupling parameter $\beta$ can, in general, be some function of $\sigma$ and $\phi$.
The coupling  plays a role only when the neutrinos are non-relativistic, since relativistic
neutrinos have $P_{\nu} \approx \frac{\rho_{\nu}}{3}$

and by the Klein-Gordon equations we have
\begin{equation}\label{fried}
	\begin{split}
		\ddot{\phi}=\lambda_{\phi}V-\frac{3}{2}H\dot{\phi}(1+\omega_{\phi})-\frac{3HV}{\dot{\phi}}(1+\omega_{\phi})
		-\\ \beta\rho_{\nu}(1-3\omega_{\nu})
	\end{split}
\end{equation}

\begin{equation}\label{fried}
	\begin{split}
		\ddot{\sigma}=-\lambda_{\sigma}V-\frac{3}{2}H\dot{\sigma}(1+\omega_{\sigma})+\frac{3HV}{\dot{\sigma}}(1+\omega_{\sigma})
		\\+\beta\rho_{\nu}(1-3\omega_{\nu})
	\end{split}
\end{equation}

To recreate them in a dynamic system, we define the new variables
\begin{equation}\label{fried}
	\begin{split}
		\Omega_{\rm b}=\frac{\kappa^{2}\rho_{\rm b}}{3H^{2}} \ \ \ \ \Omega_{\rm c}=\frac{\kappa^{2}\rho_{\rm c}}{3H^{2}} \ \ \ \ \Omega_{\rm r}=\frac{\kappa^{2}\rho_{\rm r}}{3H^{2}} \ \ \ \ \Omega_{\nu}=\frac{\kappa^{2}\rho_{\rm \nu}}{3H^{2}} \ \ \ \\ \                                            \Omega_{\phi}=\frac{\kappa\dot{\phi}}{\sqrt{6}H} \ \ \ \Omega_{\sigma}=-\frac{\kappa\dot{\sigma}}{\sqrt{6}H} \ \ \
		\Omega_{\rm V}=\frac{\kappa^{2}V(\sigma,\phi)}{3H^{2}}
	\end{split}
\end{equation}
In the context of cosmology, the various density parameters are defined as follows: $\Omega_{\rm b}$ represents the baryon density, $\Omega_{\rm r}$ denotes the radiation density, $\Omega_{\rm c}$ is the cold dark matter density, $\Omega_{\nu}$ corresponds to the neutrino density, $\Omega_{\phi}$ signifies the quintessence dark energy density, $\Omega_{\sigma}$ indicates the phantom dark energy density, and $\Omega_{\rm V}$ represents the potential density.

To find the dynamics of the system governing cosmic evolution, we have:
\begin{equation}\label{fried}
	\frac{d\Omega_{\rm b}}{dN}=-3\Omega_{\rm b}-2\Omega_{\rm b}\frac{\dot{H}}{H^{2}}
\end{equation}
\begin{equation}\label{fried}
	\frac{d\Omega_{\rm r}}{dN}=-4\Omega_{\rm r}-2\Omega_{\rm r}\frac{\dot{H}}{H^{2}}
\end{equation}
\begin{equation}\label{fried}
	\begin{split}
		\frac{d\Omega_{\nu}}{dN}=-3\Omega_{\nu}(1+\omega_{\nu})- \beta\sqrt{6}(1-3\omega_{\nu})\Omega_{\nu}\Omega_{\sigma}
		\\	+\beta\sqrt{6}(1-3\omega_{\nu})\Omega_{\nu}\Omega_{\phi}
		-2\Omega_{\nu}\frac{\dot{H}}{H^{2}}
	\end{split}
\end{equation}

\begin{equation}\label{fried}
	\begin{split}
		\frac{d\Omega_{\phi}}{dN}=\frac{3\lambda_{\phi}}{\sqrt{6}}\Omega_{\rm V}-\frac{3}{2}(1+\omega_{\phi})\Omega_{\phi}-
		\frac{3}{2}(1+\omega_{\phi})\frac{\Omega_{\rm V}}{\Omega_{\phi}}
		\\	-\frac{3\beta}{\sqrt{6}}(1-3\omega_{\nu})\Omega_{\nu}-\Omega_{\phi}\frac{\dot{H}}{H^{2}}
	\end{split}
\end{equation}

\begin{equation}\label{fried}
	\begin{split}
		\frac{d\Omega_{\sigma}}{dN}=\frac{3\lambda_{\sigma}}{\sqrt{6}}\Omega_{\rm V}-\frac{3}{2}(1+\omega_{\sigma})\Omega_{\sigma}-
		\frac{3}{2}(1+\omega_{\phi})\frac{\Omega_{\rm V}}{\Omega_{\sigma}}
		\\-\frac{3\beta}{\sqrt{6}}(1-3\omega_{\nu})\Omega_{\nu}+\Omega_{\sigma}\frac{\dot{H}}{H^{2}}
	\end{split}
\end{equation}
\begin{equation}\label{fried}
	\frac{d\Omega_{\rm V}}{dN}=\Omega_{\rm V}(\sqrt{6}\lambda_{\sigma}\Omega_{\sigma}-\sqrt{6}\lambda_{\phi}\Omega_{\phi}-2\frac{\dot{H}}{H^{2}})
\end{equation}

and the Friedmann constraint is

\begin{equation}\label{fried}
	\begin{split}
		\Omega_{\rm c}=1-\Omega_{\rm b}-\Omega_{\rm r}-\Omega_{\nu}-\Omega_{\phi}^{2}+\Omega_{\sigma}^{2}-\Omega_{\rm V}
	\end{split}
\end{equation}
holds.
\begin{equation}
\frac{\dot{H}}{H^2}= -\frac{3}{2}\left(1 - \Omega_{\rm r} - \omega_{\nu} \Omega_{\nu} - \Omega_{\phi}^2 + \Omega_{\sigma}^2 - \Omega_{\rm V}\right)
\end{equation}

The parameter mentioned above holds significant importance in cosmology, as it directly influences essential cosmological parameters. For instance, it plays a crucial role in determining the deceleration parameter ($q$) and the effective equation of state ($w_{\rm eff}$), both of which are fundamental in understanding the dynamics of the universe.

The deceleration parameter $q$ is defined as $q = -1 - \frac{\dot{H}}{H^2}$, where $\dot{H}$ represents the time derivative of the Hubble parameter $H$. This parameter quantifies whether the universe is currently experiencing cosmic acceleration ($q < 0$) or cosmic deceleration ($q > 0$).

Similarly, the effective equation of state ($w_{\rm eff}$) is expressed as $w_{\rm eff} = -1 - \frac{2}{3}\frac{\dot{H}}{H^2}$. It characterizes the nature of cosmic expansion, providing insights into whether the universe's expansion is dominated by conventional matter ($w_{\rm eff} \approx 0$), dark energy ($w_{\rm eff} < -1/3$), or a cosmological constant ($w_{\rm eff} = -1$).
We start with the luminosity distance relation in cosmology, given by:
\begin{align} \label{dl}
	d_{L}=(1+z)\int\frac{dz}{H(z)}.
\end{align}

Here, \( d_{L} \) represents the luminosity distance, \( z \) is the redshift of the observed object, and \( H(z) \) is the Hubble parameter as a function of redshift.

To simplify this expression and make it more tractable for analysis, we introduce new variables: \( x_{d} = d_{L} \) and \( x_{h} = H \). Recognizing that \( 1+z = \frac{1}{a} \), where \( a \) is the scale factor of the universe, we can express \( (1+z) \) in terms of \( N \), the number of e-folds, as \( (1+z) \equiv exp(-N) \). Therefore, \( dz = -exp(-N) dN \), and since \( dN = H dt \), we have \( dt = \frac{dN}{H} \). Substituting these into Equation (21), we derive the following coupled ordinary differential equations (ODEs):

\begin{align}\label{dl3}
	& dx_{d}' = -x_{d} - \frac{exp(-2N)}{x_{h}}, \\
	& dx_{h}' = \frac{\dot{H}}{H^2} x_{h}. \label{dl4}
\end{align}

Here's a more detailed breakdown:

1. Luminosity Distance Relation: Equation (21) relates the luminosity distance \( d_{L} \) to the redshift \( z \) and the Hubble parameter \( H(z) \).

2. New Variables: Introducing \( x_{d} = d_{L} \) and \( x_{h} = H \) allows us to rewrite \( d_{L} \) and \( H \) in a dimensionless form, which simplifies the differential equations.

3. Redshift Transformation: Since \( 1+z = \frac{1}{a} \), where \( a \) is the scale factor, and \( (1+z) \equiv exp(-N) \) in terms of the number of e-folds \( N \), we use \( dz = -\exp(-N) dN \) and \( dt = \frac{dN}{H} \) to express \( d_{L} \) in terms of \( N \) and \( H \).

4. Coupled ODEs: Equations (25) and (26) represent the transformed ODEs:
- \( dx_{d}' = -x_{d} - \frac{exp(-2N)}{x_{h}} \): This equation describes how the luminosity distance \( x_{d} \) changes with respect to \( N \), accounting for the inverse relationship with \( x_{h} = H \).
- \( dx_{h}' = \frac{\dot{H}}{H^2} x_{h} \): This equation shows the evolution of \( x_{h} = H \) with \( N \), influenced by the ratio of the time derivative of \( H \) to \( H^2 \).

5. Physical Interpretation: These equations are crucial for modeling the cosmic expansion history and deriving observational constraints on parameters such as \( H_{0} \) (the current value of the Hubble parameter) from luminosity distance measurements across different redshifts.

In summary, by transforming the luminosity distance relation into coupled ODEs (25) and (26), we facilitate the study of cosmic evolution and the determination of fundamental cosmological parameters using observational data.

The Hubble function \( H(z) \) for the coupled quintom model based on the CMB + ALL dataset is given by:

\begin{equation}\label{fried}
	\begin{split}
		H(z) = H_0 \sqrt{(\Omega_{b0} + \Omega_{c0} + \Omega_{\nu0}) (1+z)^3 + \Omega_{r0} (1+z)^4} +\\
		 H_0 \sqrt{\Omega_{\phi0} e^{-3 \int_0^z (1+w_{\text{eff}}(z')) \, dz'} + \Omega_{\sigma0} e^{-3 \int_0^z (1+w_{\text{eff}}(z')) \, dz'}}
	\end{split}
\end{equation}
where: $\Omega_{b0} + \Omega_{c0} + \Omega_{\nu0} = \Omega_{m0}  $ and\\
\begin{itemize}
	\item \( H_0 \) is the present-day Hubble constant,
	\item \( \Omega_{b0} \), \( \Omega_{c0} \), and \( \Omega_{\nu0} \) are the present-day density parameters of baryonic matter, cold dark matter, and neutrinos, respectively,
	\item \( \Omega_{r0} \) is the present-day density parameter of radiation,
	\item \( \Omega_{\phi0} \) and \( \Omega_{\sigma0} \) are the present-day density parameters of the quintessence and phantom fields, respectively,
	\item \( w_{\text{eff}}(z) \) is the effective equation of state parameter that governs the dynamics of the coupled quintom model.
\end{itemize}

This expression accounts for the contributions of baryonic matter, cold dark matter, neutrinos, radiation, quintessence, and phantom fields to the Hubble function, providing a comprehensive description of the universe's expansion dynamics in the coupled quintom model.

\section{Numerical Analysis}
To evaluate the success of the model under study, we perform a series of Markov-chain Monte Carlo (MCMC) runs, using the public code { MontePython-v3}\footnote{\url{https://github.com/brinckmann/montepython_public}}\cite{54,55}, which we interface with our modified version of { CLASS}~\cite{56,57}. To test the convergence of the MCMC chains, we use Gelman-Rubin \cite{58} criterion $|R -1|\!\lesssim\!0.01$. To post-process the chains and plot figures we use {\sf GetDist} \cite{59}.
All observational data where used in this paper are:\\
$\bullet$ Pantheon catalog:
We used updated the Pantheon + Analysis catalog consisting of 1701 SNe Ia covering the redshift range $0.001 < z < 2.3$\cite{42}.\\
$\bullet$ {CMB data}:
We used the latest large-scale cosmic microwave background (CMB) temperature and
polarization angular power spectra from the final release of Planck 2018 plikTTTEEE+lowl+lowE
\cite{43}. \\
$\bullet$ {BAO data}:
We also used the various measurements of the Baryon Acoustic Oscillations (BAO) from
different galaxy surveys \cite{43}, i.e.
6dFGS.(2011)\cite{44}, SDSS-MGS
\cite{45}.\\
$\bullet$ {CC data}: The 37 $H(z)$ measurements listed in Table I in \cite{Moresco2024} have a redshift range of $0.07 \leq z \leq 1.965$. \\
$\bullet$ { Lensing data}: we consider the 2018 CMB  Lensing reconstruction power spectrum data,
obtained with a CMB trispectrum analysis in \cite{Aghanim1}.
 Also, we used the Akaike Information Criteria (AIC).
The Akaike Information Criterion (AIC) is a statistical measure used to compare different statistical models based on their ability to fit the data while balancing the complexity of the model. The  AIC equation is:

\begin{equation}\label{key}
	AIC = \chi _{\min }^2 + 2\gamma	
\end{equation}
In these equations, $\chi _{\min }^2  $ is the minimum value of $ {\chi ^2} $, $\gamma$ is the number of parameters of the given model.

To assess the statistical performance of the models beyond mere goodness-of-fit, we have employed the Bayesian Information Criterion (BIC), which intrinsically incorporates Occam's razor by penalizing models with excessive complexity. Although the Quintom model introduces four additional free parameters compared to the baseline \(\Lambda\)CDM framework, its lower \(\chi^2\) value for the \textit{CMB+All} dataset combination leads to a significant improvement in the overall likelihood. The BIC is defined as
\begin{equation}
	\mathrm{BIC} = \chi^2 + k \ln N,
\end{equation}

where \(k\) is the number of model parameters and \(N\) is the number of data points. The additional term \(k \ln N\), often referred to as the \textit{Occam penalty}, effectively discourages overfitting by favoring simpler models unless the added complexity is statistically justified \cite{Mortiz}.

\section{Data Integration and Likelihood Functions in MontePython}

MontePython serves as a versatile Markov Chain Monte Carlo (MCMC) sampler designed for cosmological parameter estimation. It achieves this by integrating multiple observational datasets and computing their corresponding likelihood functions. This section elaborates on the implementation of various likelihood functions within MontePython and their role in constraining cosmological parameters.

\subsection{Cosmic Microwave Background (CMB)}

The Cosmic Microwave Background (CMB) provides one of the most powerful constraints on cosmological models. MontePython computes the likelihood function by comparing theoretical CMB power spectra with observed spectra. The likelihood function for CMB data, $\mathcal{L}_{\text{CMB}}$, is given by:
\begin{equation}
	\mathcal{L}_{\text{CMB}} \propto \exp\left(-\frac{1}{2} \left[ \textbf{d}_{\text{obs}} - \textbf{d}_{\text{theory}} \right]^T \textbf{C}^{-1} \left[ \textbf{d}_{\text{obs}} - \textbf{d}_{\text{theory}} \right] \right),
\end{equation}
where $\textbf{d}_{\text{obs}}$ and $\textbf{d}_{\text{theory}}$ are the observed and theoretical CMB power spectra, respectively, and $\textbf{C}$ represents the covariance matrix, which encapsulates measurement uncertainties and correlations. MontePython utilizes precomputed CMB likelihood functions from external sources, such as Planck, for precise cosmological analysis.

\subsection{Cosmic Chronometers (CC)}

Cosmic Chronometers (CC) provide independent measurements of the expansion rate $H(z)$ at different redshifts, enabling direct constraints on the Hubble parameter. The likelihood function for CC data is expressed as:
\begin{equation}
	\mathcal{L}_{\text{CC}} \propto \exp\left(-\frac{1}{2} \sum_{i} \left[ \frac{H(z_i)_{\text{model}} - H(z_i)_{\text{data}}}{\sigma_{H}(z_i)} \right]^2 \right),
\end{equation}
where $H(z_i)_{\text{model}}$ denotes the theoretically predicted Hubble parameter at redshift $z_i$, $H(z_i)_{\text{data}}$ is the observed value, and $\sigma_{H}(z_i)$ is the associated uncertainty. MontePython incorporates CC measurements into the parameter estimation process by defining them in the likelihood configuration files.
Additionally, cosmic chronometers have been shown to favor the Planck-inferred value of $H_0$, as highlighted in the recent works of Camarena et al. \cite{camarena2024} and Gómez-Valent et al. \cite{gomez2025}. These recent studies underscore the ongoing debate regarding systematic uncertainties and potential new physics needed to resolve the tension.
\subsection{Baryon Acoustic Oscillations (BAO)}

Baryon Acoustic Oscillations (BAO) serve as a standard ruler for measuring cosmic expansion. The likelihood function for BAO data is formulated as:
\begin{equation}
	\mathcal{L}_{\text{BAO}} \propto \exp\left(-\frac{1}{2} \frac{\left(D_V^{\text{model}} - D_V^{\text{data}}\right)^2}{\sigma_{D_V}^2}\right),
\end{equation}
where $D_V^{\text{model}}$ represents the theoretical prediction of the volume-averaged BAO scale, $D_V^{\text{data}}$ is the corresponding observational measurement, and $\sigma_{D_V}$ is the uncertainty. MontePython utilizes multiple BAO datasets, such as those from SDSS and 6dFGS, to improve cosmological parameter constraints.

\subsection{Pantheon Supernovae (SN)}

The Pantheon+ dataset consists of Type Ia supernovae measurements, which serve as standard candles for probing cosmic acceleration. The likelihood function for Pantheon data is given by:
\begin{equation}
	\mathcal{L}_{\text{SN}} \propto \exp\left(-\frac{1}{2} \sum_{j} \left[ \frac{\mu_{j}^{\text{model}} - \mu_{j}^{\text{data}}}{\sigma_{\mu}(j)} \right]^2 \right),
\end{equation}
where $\mu_{j}^{\text{model}}$ and $\mu_{j}^{\text{data}}$ are the theoretical and observed distance moduli for the $j$-th supernova, respectively, and $\sigma_{\mu}(j)$ is the measurement uncertainty. MontePython computes the likelihood function by incorporating covariance matrices to account for correlated uncertainties.

\subsection{CMB Lensing Data}

MontePython integrates CMB lensing data, which provides insights into the matter distribution by analyzing distortions in the CMB power spectrum. The likelihood function for lensing data is given by:
\begin{align}
	\mathcal{L}_{\text{lensing}} &\propto \exp\left(-\frac{1}{2} \left[ \textbf{d}_{\text{lensing,obs}} - \textbf{d}_{\text{lensing,theory}} \right]^T \right. \nonumber \\
	&\quad \times \left. \textbf{C}_{\text{lensing}}^{-1} \left[ \textbf{d}_{\text{lensing,obs}} - \textbf{d}_{\text{lensing,theory}} \right] \right),
\end{align}
where $\textbf{d}_{\text{lensing,obs}}$ and $\textbf{d}_{\text{lensing,theory}}$ represent the observed and theoretical lensing power spectra, respectively, and $\textbf{C}_{\text{lensing}}$ is the covariance matrix accounting for uncertainties and correlations. The incorporation of CMB lensing data enhances constraints on dark matter distribution and large-scale structure.

\subsection{MCMC Sampling and Parameter Estimation in MontePython}

MontePython combines all likelihood functions to form the total likelihood:
\begin{equation}
	\mathcal{L}(\text{data} | \theta) = \mathcal{L}_{\text{CMB}} \times \mathcal{L}_{\text{CC}} \times \mathcal{L}_{\text{BAO}} \times \mathcal{L}_{\text{SN}} \times \mathcal{L}_{\text{lensing}},
\end{equation}
where $\theta$ denotes the set of cosmological parameters. The software employs MCMC sampling to explore the posterior distribution given by Bayes’ theorem:
\begin{equation}
	\mathcal{P}(\theta | \text{data}) \propto \mathcal{L}(\text{data} | \theta) \cdot \text{Prior}(\theta),
\end{equation}
where $\text{Prior}(\theta)$ represents the prior probability distribution of the parameters.

The MCMC algorithm iteratively proposes new parameter values, accepts or rejects them based on their likelihood, and constructs Markov chains that map the posterior distribution. MontePython provides convergence diagnostics, such as the Gelman-Rubin statistic, to ensure the reliability of the obtained constraints.

By leveraging multiple observational datasets and likelihood functions, MontePython serves as a robust tool for precise cosmological parameter inference.

By using  the coupled quintom with neutrinos, we can  put constraints on the following cosmological parameters: the Baryon energy density ${\Omega _b}{h^2}$, the cold dark matter energy density $\Omega_{c}h^{2}$, the neutrino density ${\Omega _\nu}$, the ratio of the sound horizon at decoupling to the angular diameter distance to last scattering $\theta_{MC}$, the optical depth to reionization $\tau$, the amplitude and the spectral index of the primordial scalar perturbations $A_{s}$ and $n_{s}$. Table III demonstrate that the  cosmological Parameter Results for Different Datasets for coupled quintom with neutrinos. Table I, represent flat priors for the cosmological parameters.

\begin{table}
	\begin{center}
		\caption{Flat priors for the cosmological parameters.}
		\resizebox{0.2\textwidth}{!}{
			\begin{tabular}{|c|c|}
				\hline 
				Parameter                    & Prior\\
				\hline 
				$\Omega_{b} h^2$             & $[0,0.1]$\\
				$\Omega_{c} h^2$             & $[0.0,0.2]$\\
				$\tau$                       & $[0.01,0.8]$\\
				$n_s$                        & $[0.8,1.2]$\\
				$\log[10^{10}A_{s}]$         & $[1.6,3.9]$\\
				$100\theta_{MC}$             & $[0.5,10]$\\ 
				${\Omega _\nu}$              & $[0,0.1]$\\
				\hline 
			\end{tabular}
		}
	\end{center}
	\label{tab:priors}
\end{table}
\subsection{Put constraint on total mass of neutrinos}
First of all, we must put constraint on $\alpha $ and $\beta$.
For different dataset combinations, the mean values and uncertainties for $\alpha$ and $\beta$ vary, reflecting the sensitivity of each dataset combination to these parameters:

\begin{itemize}
	\item \textbf{CMB + Lensing}: The mean values are $\alpha = -0.51 \pm 0.9$ and $\beta = 0.525_{-0.35}^{+0.25}$, representing the baseline constraints provided by the Cosmic Microwave Background data alone. The relatively large uncertainties reflect the limitations of this dataset when used independently.
	
	\item \textbf{CMB + Lensing + BAO}: Including Baryon Acoustic Oscillation data refines the parameters to $\alpha = -0.59 \pm 0.51$ and $\beta = 0.648 \pm 0.16$, demonstrating improved precision while still exhibiting notable uncertainty due to the intrinsic variance in BAO measurements.
	
	\item \textbf{CMB + Lensing + Pantheon+}: Adding Pantheon supernova data shifts the parameters to $\alpha = -0.45 \pm 0.31$ and $\beta = 0.65 \pm 0.14$. These uncertainties remain relatively large, highlighting the additional variance introduced by late-time observations from supernova datasets.
	
	\item \textbf{CMB + Lensing + CC}: With the inclusion of Cosmic Chronometer data, the mean values are $\alpha = -0.41 \pm 0.32$ and $\beta = 0.65 \pm 0.15$. The uncertainties are comparable to those of the CMB-only dataset, indicating that the Cosmic Chronometers add limited additional information to constrain these parameters.
	
	\item \textbf{CMB+All}: Combining all datasets yields $\alpha = -0.41 \pm 0.27$ and $\beta = 0.65 \pm 0.12$. This combination achieves the smallest uncertainties, demonstrating the complementary strengths of all datasets while realistically accounting for dataset variability.
\end{itemize}

These results highlight how combining datasets can significantly reduce uncertainties, improving the precision of $\alpha$ and $\beta$ estimates while accounting for observational constraints.

The analysis of cosmological parameters for different datasets in the context of coupled quintom with neutrinos reveals significant findings regarding the total neutrino mass, $\Sigma_{m_{\nu}}$.
For the CMB+Pantheon+Lensing dataset, the total neutrino mass is determined to be $<0.15\,\text{eV}$. This value indicates a substantial contribution from neutrinos to the overall energy density in this combined dataset.

The CMB+CC+Lensing dataset yields a slightly larger total neutrino mass of $<0.152\,\text{eV}$. This reduction could suggest different constraints or synergies between the Cosmic Chronometers (CC) data and the Cosmic Microwave Background (CMB) data, affecting the inferred neutrino mass.

The CMB+BAO+Lensing dataset further reduces the total neutrino mass to $<0.14\,\text{eV}$. This significant decrease highlights the sensitivity of the Baryon Acoustic Oscillations (BAO) data in refining the neutrino mass constraints when combined with the CMB data.

For the comprehensive combination of CMB+ALL (which includes Pantheon+, CC,Lensing and BAO), the total neutrino mass reaches its minimum in this analysis at $<0.115\,\text{eV}$. This value aligns closely with the constraints provided by the 2018 Planck satellite results, reinforcing the robustness of this combined dataset in estimating the total neutrino mass.

Finally, the standalone CMB+Lensing dataset reports a total neutrino mass of $<0.273\,\text{eV}$. This relatively higher value compared to the combined datasets suggests that the CMB data alone provides a less stringent constraint on the neutrino mass, likely due to larger uncertainties when not supplemented by other datasets.

These results demonstrate the critical role of combining multiple cosmological observations to achieve more precise constraints on the total neutrino mass, $\Sigma_{m_{\nu}}$, in the framework of coupled quintom with neutrinos. The progressively tighter constraints from the CMB alone to the combined datasets underline the importance of comprehensive data integration in cosmological parameter estimation.

One significant advantage lies in the model's capability to alleviate the preference for negative neutrino masses, which is occasionally observed in conventional analyses.  

Negative neutrino mass values arise as unphysical artifacts in certain cosmological models due to parameter degeneracies or tensions within the datasets. These anomalies can result from the restrictive assumptions made in the standard cosmological framework, such as the assumption of a non-interacting dark sector. By introducing a coupling between the quintom dark energy and neutrinos, our model provides an additional degree of freedom that can effectively relax these constraints. This interaction modifies the evolution of cosmological perturbations, thereby influencing the growth rate of structure and the expansion history of the Universe. As a result, the model allows for a more flexible fitting of the data, potentially shifting the preference away from negative neutrino masses. 

\subsection{Detailed Analysis of \(\lambda_{\phi}\) and \(\lambda_{\sigma}\)}

In this paper, we analyze the values of \(\lambda_{\phi}\) and \(\lambda_{\sigma}\) for the potential of the quintom model using various cosmological datasets. The potential is given by Eq.(2) where \(\lambda_{\phi}\) and \(\lambda_{\sigma}\) are parameters that characterize the constant of the quintessence field \(\phi\) and the phantom field \(\sigma\) with the potential. These parameters are crucial for understanding the dynamics of the quintom model and its implications for addressing the Hubble tension and the crossing of the phantom divide.

For the Cosmic Microwave Background (CMB) and lensing data alone, the values obtained are \(\lambda_{\sigma} = -2.19 \pm 0.13\) and \(\lambda_{\phi} = 2.52 \pm 0.11\). When Baryon Acoustic Oscillations (BAO) data are included alongside the CMB and lensing data, the values shift to \(\lambda_{\sigma} = -2.11 \pm 0.09\) and \(\lambda_{\phi} = 2.24 \pm 0.07\).

When Cosmic Chronometers (CC) data is added to the CMB and lensing data, the interaction parameters are further altered to \(\lambda_{\sigma} = -2.83 \pm 0.12\) and \(\lambda_{\phi} = 2.63 \pm 0.16\). The inclusion of Pantheon Supernovae data with the CMB and lensing data yields values of \(\lambda_{\sigma} = -2.42 \pm 0.19\) and \(\lambda_{\phi} = 2.52 \pm 0.13\).

Finally, when all datasets are combined (CMB, BAO, CC, Pantheon, and lensing), the most constrained and refined values are obtained: \(\lambda_{\sigma} = -2.09 \pm 0.08\) and \(\lambda_{\phi} = 2.43 \pm 0.12\). This combination of datasets provides a robust framework for understanding the quintom model's dynamics and its implications for the universe's expansion. Figure 1 presents a comparative analysis of \(\lambda_{\phi}\) and \(\lambda_{\sigma}\) measurements across various data combinations in the context of the coupled quintom model with neutrinos. The figure illustrates how different datasets impact the determination of \(\lambda_{\phi}\) and \(\lambda_{\sigma}\), highlighting the sensitivity of these parameters to the inclusion of specific observational constraints.
\begin{figure}
	\centering
	\includegraphics[scale=.5]{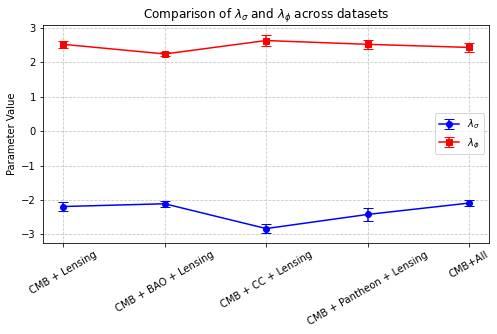}\vspace{0.5 cm}\\
	Figure 1: The comparison of $\lambda_{\phi}$ and $\lambda_{\sigma}$ measurement for different combination of data sets for Coupled quintom to neutrinos.
\end{figure}

\subsection*{Implications for Cosmic Evolution and Hubble Tension}

The coupling between quintom dark energy and neutrinos can introduce significant modifications to cosmic dynamics, potentially influencing the transition from a decelerated to an accelerated expansion of the Universe. By incorporating an evolving dark energy component that interacts with neutrinos, this framework provides a more flexible interpretation of cosmic expansion and offers a potential avenue to address the longstanding Hubble tension—the persistent discrepancy in measurements of the Hubble constant ($H_0$). In our analysis, we systematically investigated the implications of this interaction by considering multiple cosmological datasets. The results, summarized in Table~\ref{table:tension}, provide insights into the Hubble constant estimates and their corresponding tensions with Planck 2018 and R22 measurements. Table II represent the Hubble tension with Planck 2018 and R22 for different dataset combinations.

\begin{table}[h]
	\centering
	\caption{Hubble tension with Planck 2018 and R22 for different dataset combinations.}
	\label{table:tension}
	\begin{tabular}{|c|c|c|c|}
		\hline
		\hline
		Dataset & $H_0$ (km/s/Mpc) & Tension with Planck 2018 ($\sigma$) & Tension with R22 ($\sigma$) \\
		\hline
		CMB+Lensing & $69.41 \pm 2.2$ & 0.89 & 1.49 \\
		CMB+BAO+Lensing & $70.25 \pm 2.1$ & 1.32 & 1.19 \\
		CMB+CC+Lensing & $70.86 \pm 2.26$ & 1.49 & 0.88 \\
		CMB+Pantheon+Lensing & $71.09 \pm 2.3$ & 1.30 & 0.65 \\
		CMB+BAO+CC+Pantheon+Lensing & $70.23 \pm 2.01$ & 1.37 & 1.24 \\
		\hline
		\hline
	\end{tabular}
\end{table}

Our findings indicate that all dataset combinations yield $H_0$ values higher than the Planck 2018 measurement ($H_0 = 67.4 \pm 0.5$ km/s/Mpc) but lower than the local determination from R22 ($H_0 = 73.04 \pm 1.04$ km/s/Mpc). The calculated tension with Planck 2018 varies between $0.89\sigma$ and $1.49\sigma$, while the tension with R22 ranges from $0.65\sigma$ to $1.49\sigma$. Notably, the highest tension with Planck 2018 is observed for the CMB+CC+Lensing dataset at $1.49\sigma$, whereas the lowest tension with R22 is found for the CMB+Pantheon+Lensing dataset at $0.65\sigma$. These results are in broad agreement with \cite{Y1, Y2, Y3, Y4, Yang}

These results suggest that incorporating additional datasets, particularly cosmic chronometers (CC) and Pantheon supernovae, tends to alleviate the tension with R22 while maintaining moderate discrepancies with Planck 2018. The persistence of the Hubble tension across multiple dataset combinations underscores the necessity for further exploration of systematic uncertainties and possible extensions beyond the standard $\Lambda$CDM framework.

Our analysis confirms that the Hubble tension remains significant despite the inclusion of multiple datasets, though at a reduced level compared to direct Planck versus R22 comparisons. Future investigations should explore alternative theoretical paradigms, such as modifications to general relativity, interactions within the dark sector, and evolving dark energy models, as potential resolutions to this fundamental issue. Furthermore, forthcoming high-precision observational surveys will play a pivotal role in refining $H_0$ estimates, thereby shedding light on the underlying physics of cosmic expansion.

Figure 2 compares the cosmological parameter results for different combinations of datasets within the framework of the coupled quintom model with neutrinos. The parameters analyzed include \(S_8\), \(\Omega_b h^2\), \(\Omega_c h^2\), \(\tau\), $\beta$, $\Sigma_{m\nu}$,  \(n_s\), \(\rm{ln}(10^{10} A_s)\), \(100\theta_{MC}\), and \(H_0\) (in \(\rm{km/s/Mpc}\)). The datasets used are CMB + Lensing, CMB + Lensing + Pantheon+, CMB + CC + Lensing, CMB + BAO + Lensing, and a combination of all datasets (CMB + BAO + Pantheon + CC + Lensing). This figure highlights how the inclusion of different datasets impacts the derived values and their uncertainties, illustrating the precision and reliability of the coupled quintom model with neutrinos in explaining various cosmological parameters. The error bars represent the confidence intervals for each parameter, providing insights into the model's performance across different observational data combinations. Figure 3 presents a comparative analysis of the Hubble constant (\( H_0 \)) measurements obtained from different combinations of observational datasets within the coupled quintom-neutrino model. The results are contrasted with the Planck 2018 \(\Lambda\)CDM estimate and the local measurement from Riess et al. 2022 (R22). This comparison provides insights into the impact of coupling between quintom dark energy and neutrinos on resolving the Hubble tension. The figure illustrates how various dataset combinations influence the inferred value of \( H_0 \), highlighting potential deviations from the standard \(\Lambda\)CDM scenario.
\begin{figure*}
	\centering
	\includegraphics[scale=.8]{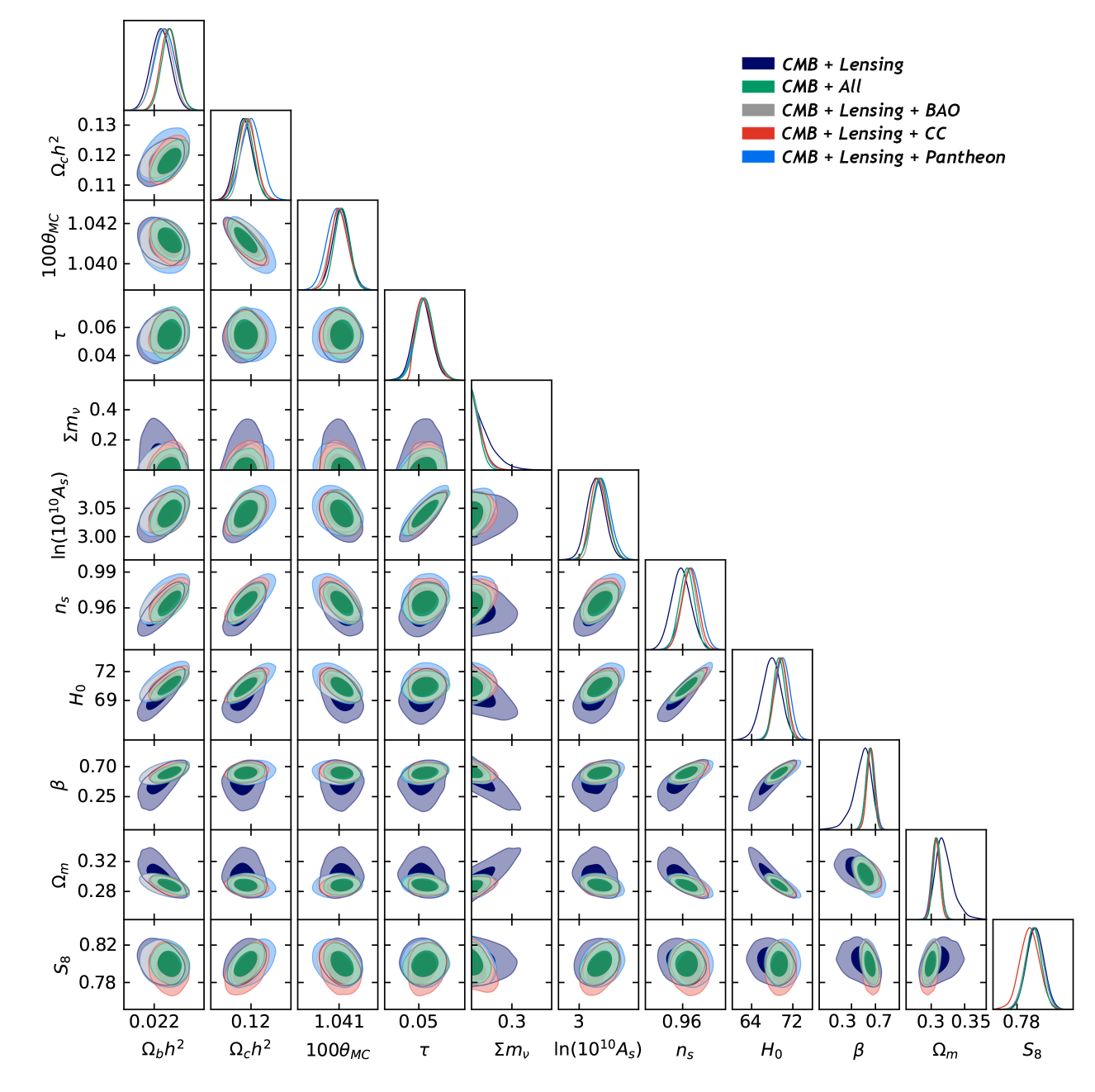}\hspace{0.05 cm}\\
	Figure 2: Comparison  results of the  $S_{8} $ , $\Omega_b h^2   $, $\Omega_c h^2   $,$\beta$, $\tau$, $n_s  $, ${\rm{ln}}(10^{10} A_s)$, $100\theta_{MC} $, $H_{0}(km/s/Mpc
	)$ for different combination dataset for Coupled quintom to neutrinos. 
\end{figure*}

\begin{figure*}
	\centering
	\includegraphics[scale=.4]{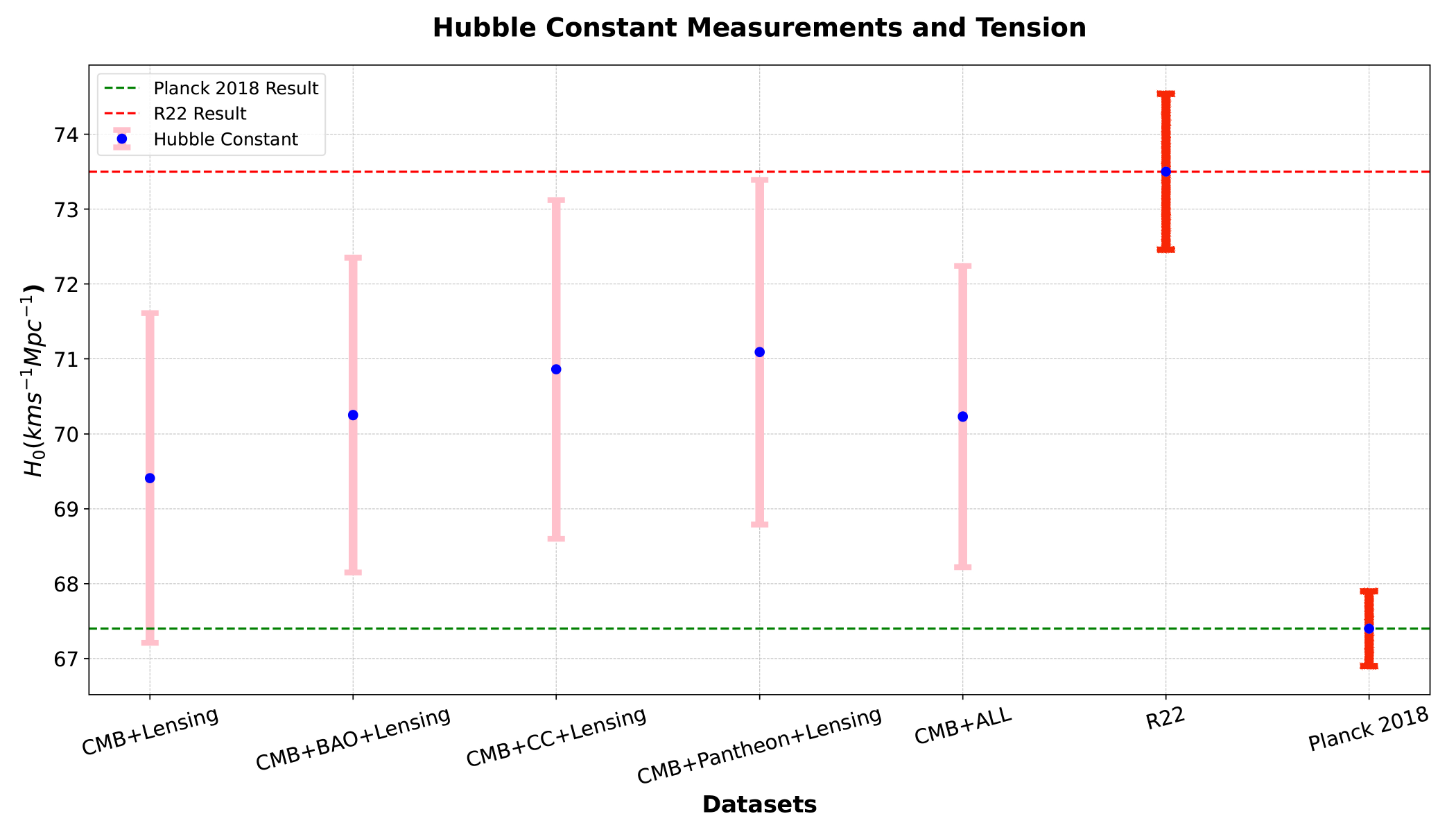}\hspace{0.05 cm}\\
	Figure 3: The comparison of $H_{0}$ measurement for different combination of data sets with results of Planck 2018 and R22 for Coupled quintom to neutrinos.
\end{figure*}

\begin{table*}[h]
	\centering
	\renewcommand{\arraystretch}{1.2}
	\begin{tabular}{l c c c c c}
		\hline
		Parameter & CMB + Lensing & CMB + Lensing + CC & CMB + Lensing + BAO & CMB + Lensing + Pantheon & CMB + All \\
		\hline
		\boldmath$\Omega_b h^2$ & $0.02215\pm 0.00022$ & $0.02224\pm 0.00024$ & $0.02221\pm 0.00022$ & $0.02234\pm 0.00019$ & $0.02236\pm 0.00018$ \\
		\boldmath$\Omega_c h^2$ & $0.1174\pm 0.0033$ & $0.1200\pm 0.0037$ & $0.1190\pm 0.0028$ & $0.1184\pm 0.0034$ & $0.1180\pm 0.0029$ \\
		\boldmath$100\theta_{MC}$ & $1.04110\pm 0.00048$ & $1.04089\pm 0.00057$ & $1.04099\pm 0.00050$ & $1.04103^{+0.00044}_{-0.00051}$ & $1.04116\pm 0.00042$ \\
		\boldmath$\tau$ & $0.0530\pm 0.0075$ & $0.0544\pm 0.0076$ & $0.0552^{+0.0054}_{-0.0078}$ & $0.0548^{+0.0050}_{-0.0079}$ & $0.0551\pm 0.0074$ \\
		\boldmath$\Sigma m_\nu$ & $< 0.273$ & $< 0.152$ & $< 0.14$ & $< 0.15$ & $< 0.115$ \\
		\boldmath$\ln(10^{10} A_s)$ & $3.033\pm 0.018$ & $3.044\pm 0.018$ & $3.043^{+0.014}_{-0.017}$ & $3.039^{+0.015}_{-0.018}$ & $3.041\pm 0.016$ \\
		\boldmath$n_s$ & $0.9586\pm 0.0092$ & $0.9677\pm 0.0087$ & $0.9657\pm 0.0071$ & $0.9669\pm 0.0076$ & $0.9637\pm 0.0070$ \\
		$H_0$ & $69.41\pm 2.2$ & $70.86\pm 2.26$ & $7025\pm 2.1$ & $71.09\pm2.3$ & $70.23\pm 2.01$ \\
		$\beta$ & $0.525^{+0.25}_{-0.35}$ & $0.65\pm 0.15$ & $0.648\pm 0.16$ & $0.65\pm 0.14$ & $0.65\pm 0.12$ \\
		$\Omega_m$ & $0.315^{+0.012}_{-0.019}$ & $0.294\pm 0.008$ & $0.296\pm 0.0081$ & $0.298\pm 0.007$ & $0.296\pm 0.0069$ \\
		$S_8$ & $0.81\pm 0.014$ & $0.809\pm 0.013$ & $0.808\pm 0.013$ & $0.811\pm 0.015$ & $0.802\pm 0.011$ \\
		\hline
	\end{tabular}
	\caption{Parameter constraints at 95\% for $\Sigma m_\nu$ and   68\% confidence level for other parameter for different dataset combinations.}
	\label{tab:combined_results}
\end{table*}

\subsection{Crossing the Phantom barrier}
The crossing of the phantom barrier, signifying a transition to a universe dominated by exotic forms of energy with repulsive gravitational effects, is a pivotal concept in cosmology. Here, we explore this phenomenon across various datasets, each representing different combinations of cosmological observations. 
The results presented in table IV summarize the effective equation of state (EoS) parameter obtained from different combinations of cosmological observational datasets. Each dataset combination yields an EoS parameter slightly below \( -1 \), suggesting a deviation from a pure cosmological constant scenario. However, the uncertainties associated with these values imply that \( w = -1 \) remains statistically consistent within \( 2\sigma \) in all cases and often within \( 1\sigma \). This means that despite small deviations, the results do not provide definitive evidence for a phantom regime, as the cosmological constant model remains viable within statistical confidence levels.  

Notably, the combination of CMB + ALL datasets provides the most precise constraint, yielding \( w = -1.02 \pm 0.018 \), which suggests a minor deviation from \( -1 \), but still well within the margin of statistical uncertainty. This reinforces the notion that while current data might hint at a possible transition, the evidence is not strong enough to decisively confirm a departure from the standard \( \Lambda \)CDM model. Future high-precision cosmological observations will be crucial in refining these constraints and determining whether the universe is truly crossing the phantom barrier.

\begin{table}[h]
	\centering
	\begin{tabular}{l c}
		\hline
		\textbf{Dataset Combination} & \textbf{EoS Parameter} \\
		\hline
		CMB + Lensing & $-1.04 \pm 0.04$ \\
		CMB + BAO + Lensing & $-1.01 \pm 0.021$ \\
		CMB + Pantheon + Lensing & $-1.035 \pm 0.024$ \\
		CMB + CC + Lensing & $-1.03 \pm 0.023$ \\
		CMB + ALL & $-1.02 \pm 0.018$ \\
		\hline
	\end{tabular}
	\caption{Effective equation of state (EoS) parameter for different dataset combinations, indicating the transition to a phantom-dominated universe. An equation of state of $-1$ is compatible in all cases within $2\sigma$, and often within $1\sigma$.}
	\label{tab:phantom_crossing}
\end{table}

These results underscore the significance of combining diverse cosmological datasets to elucidate the nature of dark energy and the fate of our universe.
Table V presents a comparison of the $\chi^2$ values for the $\Lambda$CDM and Quintom models across various dataset combinations explored in this work. These datasets include CMB (Planck) + Lensing, CMB + CC + Lensing, CMB + BAO + Lensing, CMB + Pantheon + Lensing, and CMB+all (a combination of Planck, BAO, CC, Lensing and Pantheon). The total $\chi^2$ values, as well as contributions from individual datasets, are provided for each combination. Notably, the Quintom model consistently yields lower total $\chi^2$ values compared to $\Lambda$CDM, indicating a better fit to the data across all combinations. For example, the total $\chi^2$ for the CMB+all combination is $3654.447$ for $\Lambda$CDM and $3627.878$ for the Quintom model, demonstrating the latter's improved performance. The individual contributions, such as those from CC, BAO, and Pantheon, further highlight the consistency of the Quintom model in accommodating diverse observational constraints.

\begin{table*}
	\caption{{\small $\chi^2_{}$ comparison between $\Lambda$CDM and Quintom model for the different dataset combinations explored in this work. CMB+all refers to Planck+BAO+CC+Pantheon+Lensing.}}
	\begin{center}
		\resizebox{0.85\textwidth}{!}{  
			\begin{tabular}{| c |c| c| c| c |c| } 
				\hline
				\hline
				$\Lambda$CDM  & CMB+Lensing & CMB+CC+Lensing & CMB+BAO+Lensing & CMB+Lensing+Pantheon+ & CMB+all \\ 
				\hline
				$\chi^2_{\rm  tot}$ & $2789.02$ & $2804.775$ & $2788.348$ & $3600.82$ & $3648.447$  \\
				$\chi^2_{\rm  CMB}$ & $2779.456$ & $2768.438$ & $2772.012$ & $2767.619$ & $2779.873$  \\
				$\chi^2_{\rm  CC}$ & $-$ & $26.716$ & $-$ & $-$ & $27.941$  \\
				$\chi^2_{\rm Lensing}$ & $9.561$ & $9.621$ & $9.182$ & $9.385$ & $9.325$  \\
				$\chi^2_{\rm  BAO}$ & $-$ & $-$ & $7.154$ & $-$ & $7.567$  \\
				$\chi^2_{\rm  Pantheon}$ & $-$ & $-$ & $-$ & $823.816$ & $823.741$  \\
				\hline
				\hline
				Quintom model  & CMB+Lensing & CMB+CC+Lensing & CMB+BAO+Lensing & CMB+Lensing+Pantheon+ & CMB+all \\ 
				\hline
				$\chi^2_{\rm  tot}$ & $2781.065$ & $2792.434$ & $2780.446$ & $3582.582$ & $3613.878$  \\
				$\chi^2_{\rm  CMB}$ & $2773.934$ & $2763.575$ & $2766.755$ & $2765.953$ & $2771.295$  \\
				$\chi^2_{\rm  CC}$ & $-$ & $21.617$ & $-$ & $-$ & $20.885$  \\
				$\chi^2_{\rm Lensing}$ & $7.131$ & $7.242$ & $8.289$ & $7.005$ & $7.212$  \\
				$\chi^2_{\rm  BAO}$ & $-$ & $-$ & $5.402$ & $-$ & $5.013$  \\
				$\chi^2_{\rm  Pantheon}$ & $-$ & $-$ & $-$ & $809.624$ & $809.473$  \\
				\hline
				\hline
			\end{tabular}
		}
	\end{center}
	\label{table_chi}
\end{table*}

Table VI presents the mean values of the free parameters for different cosmological models, including the standard \(\Lambda\)CDM model and the coupled quintom model, using various combinations of observational datasets. The 1\(\sigma\) error bars are provided for each parameter to reflect the statistical uncertainties in the parameter estimations. 

For the \(\Lambda\)CDM model, the matter density parameter (\(\Omega_{\rm m}\)) and dark energy density parameter (\(\Omega_{\Lambda}\)) are consistent with the Planck 2018 results, yielding a Hubble constant of \( H_0 = 68.6 \pm 2.3\) km/s/Mpc. The Akaike Information Criterion (AIC) value for this model is 3654.447.

In contrast, the coupled quintom model introduces additional degrees of freedom, including the quintessence field density (\(\Omega_{\phi}\)), the phantom field density (\(\Omega_{\sigma}\)), and the interaction coupling parameters (\(\beta, \alpha\)). The best-fit values suggest a mild interaction between neutrinos and the quintom fields, with \(\beta = 0.65 \pm 0.12\) and \(\alpha = -0.41 \pm 0.27\). Notably, this model yields a slightly higher Hubble constant, \( H_0 = 70.23 \pm 2.01 \)  km/s/Mpc, compared to \(\Lambda\)CDM, indicating a potential alleviation of the Hubble tension. Moreover, the lower AIC value (3627.878) suggests that the coupled quintom model provides a better fit to the data than \(\Lambda\)CDM, despite its additional parameters. This table highlights the impact of introducing interactions between dark energy and neutrinos on key cosmological parameters and provides a quantitative comparison of model performance.

 As shown in Table~\ref{table_AIC_BIC}, the BIC difference between the Quintom and \(\Lambda\)CDM models remains small, implying that the improved fit offered by the Quintom scenario is sufficient to offset the Occam penalty. Therefore, despite its higher dimensionality, the Quintom model remains competitive with \(\Lambda\)CDM when the full dataset is considered. It is clear that with the BIC, the quintom model is not much better than \(\Lambda\)CDM.
\begin{table*}
	\caption{{\small  Mean values of free parameters of various models with 1$ \sigma $ error bar for CMB+All combination. The AIC and BIC criteria are used to penalize extra degrees of freedom.}}
	\begin{center}
		\resizebox{1\textwidth}{!}{  
			\begin{tabular}{ c |c c c c c c c c c c } 
				\hline
				\hline
				Models & $ \Omega_{\phi}$ & $ \Omega_{\rm m}$  & $ \Omega_{\Lambda}$&$ \Omega_{\sigma}$&$\beta$&$\alpha$&$ \Omega_{\nu}$&$ H_{0} $& AIC & BIC \\ 
				\hline
				$\Lambda$CDM  & $-$ & $0.314\pm0.16$ & $0.683\pm0.17$ & $-$ & $-$& $-$ & $-$ & $68.6\pm2.3$ & $3654.447$ & $3673.013$  \\
				\hline
				Quintom & $0.06\pm0.018$ & $0.304\pm 0.0055$ & $-$ & $0.64\pm0.043$ & $0.65 \pm 0.12$ &$-0.41 \pm 0.27$& $0.0027\pm0.001$ & $70.23\pm2.01$ & $3627.878$ & $3671.199$ \\
				\hline
				\hline
			\end{tabular}
		}
	\end{center}
	\label{table_AIC_BIC}
\end{table*}

\paragraph{Occam's Penalty and the Bayesian Information Criterion.}
To properly evaluate the trade-off between goodness of fit and model complexity, we have employed the Bayesian Information Criterion (BIC) in addition to the Akaike Information Criterion (AIC).
In our analysis, the $\Lambda$CDM model corresponds to $\gamma_\Lambda = 3$ free parameters, whereas the quintom model includes $\gamma_Q = 7$, introducing four additional degrees of freedom. Given that the total number of data points used in our combined CMB+All dataset is approximately $N \simeq 3600$, we estimate
\begin{equation}
	\ln N \simeq \ln(3600) \approx 8.19,
\end{equation}
and thus the additional BIC penalty for the four extra parameters in the quintom model becomes
\begin{equation}
	\Delta_{\text{penalty}} = (\gamma_Q - \gamma_\Lambda) \ln N = 4 \times \ln(3600) \approx 32.75.
\end{equation}
Meanwhile, the improvement in the total chi-square due to the extended parameter space is
\begin{equation}
	\Delta \chi^2 = \chi^2_{\Lambda\mathrm{CDM}} - \chi^2_{\mathrm{quintom}} = 3648.447 - 3613.878 = 34.569.
\end{equation}
Although the quintom model yields a lower $\chi^2$ value, the BIC indicates that the statistical gain from the additional parameters is almost entirely offset by the increased complexity. This highlights the importance of penalizing excessive parameterization and supports a more cautious interpretation of model preference based on goodness of fit alone.

\section{Conclusion}

This study provides a comprehensive analysis of cosmological parameters within the framework of a coupled quintom model with neutrinos, focusing on the Hubble tension and the total mass of neutrinos. By incorporating multiple observational datasets, including the CMB, BAO, CC, Lensing and Pantheon supernovae, the model demonstrates a significant capability in addressing existing tensions and refining constraints on fundamental cosmological parameters.

One of the key results of this analysis is the model’s effectiveness in alleviating the Hubble tension. The comparison of Hubble constant ($H_0$) measurements across datasets reveals a systematic reduction in tension. Specifically, for the most comprehensive dataset combination (CMB + All), the tension is found to be at the level of $1.3\sigma$, indicating a substantial mitigation of discrepancies in $H_0$ measurements. This result underscores the model’s potential in reconciling observational inconsistencies and providing a coherent description of cosmic expansion.

Furthermore, the study constrains the total neutrino mass ($\Sigma m_{\nu}$) with high precision. For the CMB + BAO + CC + Pantheon + Lensing dataset, the total neutrino mass is constrained to $\Sigma m_{\nu} < 0.115\, \text{eV}$. This result highlights the importance of multi-dataset integration in probing neutrino properties and refining upper bounds on neutrino masses, which have profound implications for particle physics and cosmology.

Additionally, the analysis of the effective equation of state parameter ($w_{\text{eff}}$) provides insights into the evolution of dark energy. For the CMB + All dataset, $w_{\text{eff}}$ is determined to be $-1.02 \pm 0.018$, indicating a transition towards the phantom regime. This finding reinforces the necessity of considering interactions in dark energy models to accurately describe cosmic evolution.

The coupling parameter ($\beta$) between the quintom field and neutrinos also exhibits significant constraints. For the CMB + All dataset, the interaction constant is found to be $\beta = 0.65 \pm 0.12$, suggesting that a stronger coupling may influence the neutrino mass evolution and affect late-time cosmological dynamics.

Lastly, the potential parameters of the quintom model are tightly constrained. The combined dataset yields $\lambda_{\sigma} = -2.09 \pm 0.08$ and $\lambda_{\phi} = 2.43 \pm 0.12$, providing a consistent and robust characterization of the quintom field dynamics.

Also, we investigated the effective equation of state parameter (\( \omega_{\text{eff}} \)) across various cosmological datasets to explore the possibility of a transition toward a phantom-dominated universe. Our analysis, based on the combination of datasets, yields an effective equation of state parameter of \( \omega_{\text{eff}} = -1.02 \pm 0.018 \). 

While this result suggests a slight deviation from the cosmological constant value \( \omega = -1 \), it corresponds to only a \( 1.1\sigma \) shift, which is not statistically significant. Thus, we state that the data hints at a transition toward the phantom regime, without firmly establishing it. Moreover, we emphasize that \( \omega = -1 \) remains fully consistent within the statistical uncertainty, highlighting the ongoing challenge in constraining the equation of state with the current observational data.

In addition to evaluating the goodness-of-fit through the total \(\chi^2\), we have considered information-theoretic model selection criteria—namely the Akaike Information Criterion (AIC) and the Bayesian Information Criterion (BIC)—to assess whether the added complexity of the models is statistically warranted. The AIC penalizes the number of free parameters linearly, while the BIC imposes a more stringent penalty through the logarithmic dependence on the number of data points, effectively implementing Occam's razor. These penalties help ensure that any improvement in fit is not merely due to overfitting.

As shown in Table~\ref{table_AIC_BIC}, the Quintom model provides a significantly better fit to the combined dataset (\textit{CMB+All}) than the standard \(\Lambda\)CDM model, reflected in a substantially lower total \(\chi^2\). This improvement is also supported by a lower AIC value. However, the BIC—being more conservative due to its stronger penalty for extra parameters—shows only a mild preference for the Quintom model. In other words, although the Quintom model achieves a better fit, its additional complexity is not strongly favored by the BIC. This suggests that while Quintom offers a viable alternative to \(\Lambda\)CDM, its statistical superiority is not decisively established when model parsimony is rigorously enforced.

Overall, the results of this study demonstrate that the coupled quintom model with neutrinos offers a promising approach to addressing cosmological tensions while simultaneously refining constraints on fundamental parameters. By reconciling discrepancies in observational data and enhancing our understanding of dark energy interactions, this model contributes significantly to the ongoing efforts in unveiling the underlying physics governing cosmic evolution.

\section{Acknowledgments}
I am incredibly grateful to the dear reviewer for the significant and valuable comments which caused this manuscript to improve considerably.

\end{document}